\documentclass[conference, a4paper]{IEEEtran}
\IEEEoverridecommandlockouts
\usepackage{cite}
\usepackage{amsmath,amssymb,amsfonts, multirow, textcomp}
\usepackage{algorithmic}
\usepackage{graphicx}
\usepackage{textcomp, booktabs}
\usepackage{xcolor}
\def\BibTeX{{\rm B\kern-.05em{\sc i\kern-.025em b}\kern-.08em
    T\kern-.1667em\lower.7ex\hbox{E}\kern-.125emX}}
\usepackage[latin9]{inputenc}    
    
\makeatletter

\def\ps@IEEEtitlepagestyle{
  \def\@oddfoot{\mycopyrightnotice}
  \def\@evenfoot{}
}
\def\mycopyrightnotice{
  {\footnotesize \copyright~2019 IEEE, Accepted for publication in the Proceedings of IEEE WCNC 2019 Workshops.\hfill} 
  \gdef\mycopyrightnotice{}
}

\begin{document}

\title{On the Resource Utilization of Multi-Connectivity Transmission for URLLC Services in 5G New Radio\\
\thanks{This work has partly been performed in the framework of the Horizon $2020$ project ONE-5G (ICT$-760809$) receiving funds from the European Union, and partly under Academy of Finland 6Genesis Flagship (grant no. $318927$). The authors would like to acknowledge the contributions of their colleagues in the project, although the views expressed in this work are those of the authors and do not necessarily represent the project.
\newline
This work was carried out while N.H. Mahmood was at Aalborg University, Denmark.
}
}

\author{
Nurul Huda Mahmood{\small $^{1}$},
Ali Karimi{\small $^{2}$},
Gilberto Berardinelli{\small $^{2}$},
Klaus I. Pedersen{\small $^{2,3}$} and 
Daniela Laselva{\small $^{3}$}\\
\fontsize{9}{9}\selectfont\itshape
$~^{1}$Centre for Wireless Communications, University of Oulu, Finland.\\
$~^{2}$Department of Electronic Systems, Aalborg University, Denmark.\\
$~^{3}$Nokia Bell Labs, Aalborg, Denmark.\\
$~^{}$Emails: NurulHuda.Mahmood@oulu.fi, \{alk, gb\}@es.aau.dk, \{\textit{firstname.lastname}\}@nokia.com\\
}

\maketitle

\begin{abstract}
Multi-connectivity with packet duplication, where the same data packet is duplicated and transmitted from multiple transmitters, is proposed in 5G New Radio as a reliability enhancement feature. This paper presents an analytical study of the outage probability enhancement with multi-connectivity, and analyses its cost in terms of resource usage. The performance analysis is further compared against conventional single-connectivity transmission. Our analysis shows that, for transmission with a given block error rate target, multi-connectivity results in more than an order of magnitude outage probability improvement over the baseline single-connectivity scheme. However, such gains are achieved at the cost of almost doubling the amount of radio resources used. Multi-connectivity should thus be selectively used such that its benefits can be harnessed for critical users, while the price to pay in terms of resource utilization is simultaneously minimized. 
\end{abstract}

\begin{IEEEkeywords}
Dual-connectivity/multi-connectivity, 5G, new radio, URLLC, PDCP duplication.
\end{IEEEkeywords}

\section{Introduction}
\label{sec:introduction}
The fifth generation (5G) New Radio (NR) wireless network standard introduced Ultra Reliable Low Latency Communication (URLLC) service class with the design goals of providing very high reliability and low latency wireless connectivity~\cite{3gppTS38300}. Its use cases include Industry 4.0 automation, communication for intelligent transport services and tactile Internet. Different design goals have been identified for different applications, with one of the more stringent targets being a $99.999\%$ reliability (i.e. $10^{-5}$ outage probability) at a maximum one way data-plane latency of \textit{one} millisecond (ms).

A number of solutions addressing the scheduling and resource allocation aspects of URLLC have been proposed in the literature.  These include short transmission time intervals (TTI)\cite{3gppTR38912}, faster processing~\cite{pedersenAgileScheduler_comMag2018} and enhanced URLLC-aware scheduling solution including pre-emption~\cite{karimi_ieeeAccess2018}. 

On the other hand, examples of studies investigating URLLC from an analytical perspective include~\cite{bennis_tail2018, berardinelli_reliabilityAnalysis_ieeeAccess2018}´, among others. Reference~\cite{bennis_tail2018} breaks down URLLC into three major building blocks, namely: \textit{(i)} \textit{risk} representing decision making uncertainty, \textit{(ii)} extreme values at the \textit{tail} of a distribution affecting high reliability, and \textit{(iii)} the \textit{scale} at which various network elements requiring URLLC services are deployed. The authors then discuss various enablers of URLLC and their inherent tradeoffs, and present several mathematical tools and techniques that can be used to design URLLC solutions. Reference~\cite{berardinelli_reliabilityAnalysis_ieeeAccess2018} analyse the reliability of uplink grant-free schemes, which have the potential of reducing the latency by avoiding the handshaking procedure for acquiring a dedicated scheduling grant, and demonstrate their latency benefits with respect to conventional grant-based approach. 

The stochastic nature of the wireless channel is one of the main constraints in achieving the stringent URLLC service requirements. Ensuring high reliability requires overcoming variations in the received signal strength caused by the channel. Diversity is a well proven technique in this regard~\cite{tseFundamentalsBook}. It is now being revisited as a reliability improvement feature for URLLC services through Multi-Channel Access (MCA) solutions~\cite{mahmood_mca_comMag2018, 3gppTS37340, MLL+18_DC}. MCA is a promising family of radio resource management approach that allows a user equipment (UE) to be simultaneously served over multiple channels through one or more transmitting nodes. Carrier aggregation is an example of single-node MCA, whereas examples of multi-node MCA include joint transmission, multi-connectivity (MC) and downlink-uplink decoupling~\cite{mahmood_mca_comMag2018}.  

This paper specifically addresses reliability oriented MC with packet duplication, focusing on the downlink transmission direction. MC is a generalization of the dual-connectivity (DC) concept, first standardized in 3GPP release-12 as a throughput enhancement feature~\cite{rosa_dc_comMag2016}. MC with packet duplication involves duplication of a packet destined for a particular UE, which is then transmitted to the UE through multiple transmitting nodes. The current 5G NR release-15 standard specifies that packet duplication has to occur at the Packet Data Convergence Protocol (PDCP) layer~\cite{3gppTS38300}. Transmissions from the individual nodes are independent at the lower layers, and thus can be transmitted with different transmission parameters, e.g., modulation and coding schemes (MCS). Reliability improvement with MC introduces transmission diversity that can overcome some of the causes of transmission failures, such as deep fades and/or strong interference.

Data duplication greatly enhances the probability of successfully receiving the data packet, albeit at the cost of increased resource usage. Recently, the 3GPP has acknowledged the resource efficiency challenge of PDCP duplication and is studying how to improve the packet delivery efficiency in future releases~\cite{3gppTR38825}. Several solutions have been discussed, comprising selective duplication to minimize the used resources for duplicates, and timely discarding of redundant duplicates. These enhancements are seen even more crucial when envisioning the extension of the number of radio links participating to the packet delivery or to the number of simultaneous duplicates as compare to release-15.

This paper provides a thorough analysis of MC. In particular, the reliability improvement with MC, measured in terms of the outage probability gain, is analytically derived. In addition, the operational cost of MC in terms of resource utilization is also analysed. Due to the limited space, a detailed analysis of the transmission latency could not be included in this work. The latency aspect of URLLC is implicitly covered by the considered 5G numerology, which allows at most a single hybrid automatic repeat request (HARQ) retransmission within the assumed one ms latency budget. 

The rest of the paper is organized as follows. The system model is introduced in Section~\ref{sec:systemModel}. Section~\ref{sec:reliabilitEnhancementWithMC} and Section~\ref{sec:resourceUsage} present a detailed analysis of the outage probability enhancement and the corresponding resource usage with baseline SC and reliability-oriented MC, respectively. Numerical results are then presented in Section~\ref{sec:results} followed by concluding remarks in Section~\ref{sec:conclusion}.


\section{Setting the Scene}
\label{sec:systemModel}


\subsection{Latency Components}
\begin{figure}[htb]
    \centering
    \includegraphics[width=0.99\columnwidth]{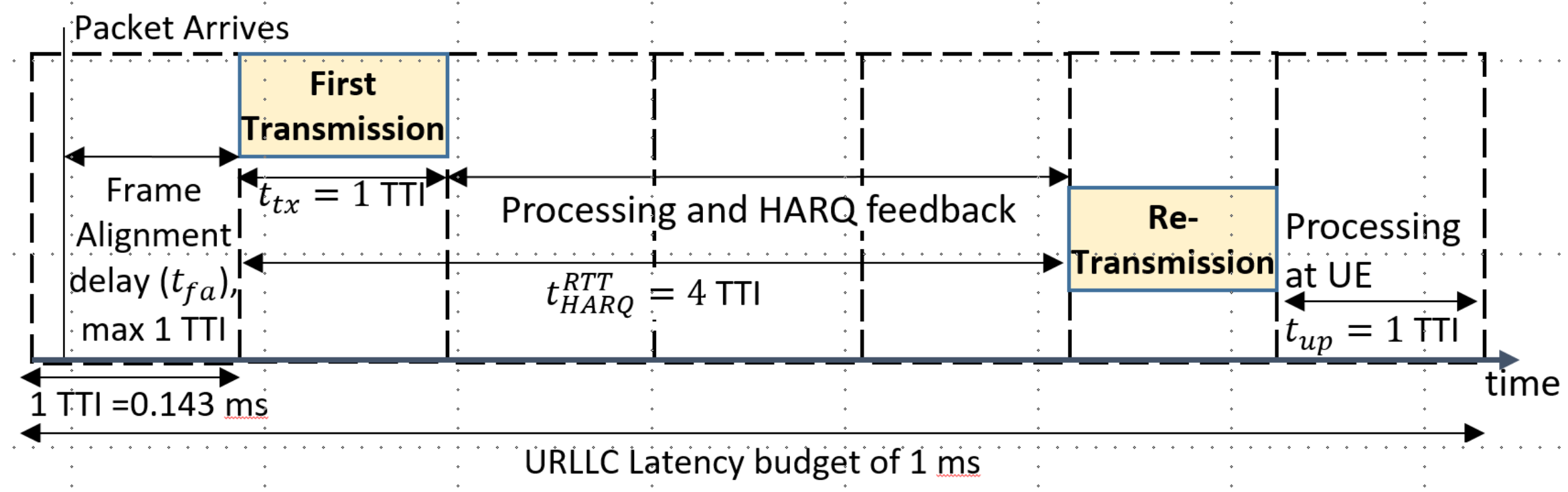}
    \caption{URLLC latency budget of one ms can accommodate maximum one HARQ retransmission at four OFDM symbols mini-slot with $30$ kHz sub-carrier spacing, corresponding to $0.143$ ms TTI.}
    \label{fig:processingTime}
\end{figure}

The downlink one-way latency of a given transmission $(\varUpsilon)$ is defined from the time a payload arrives at the lower layer of the transmitting base station (BS), until it is successfully decoded at the UE. If the UE correctly decodes the packet in the first transmission, the latency is that of a single transmission, as illustrated in Figure~\ref{fig:processingTime}, and is given by
\begin{equation}
\varUpsilon= t_{fa}+ t_{bp}+ t_{tx}+ t_{up},	
\end{equation} 
where $t_{fa}$ is the frame alignment delay. The payload transmission time is denoted by $t_{tx}$. The processing times at the BS and the UE are represented by $t_{bp}$ and $t_{up}$, respectively. 

The frame alignment delay is a random variable uniformly distributed between \textit{zero} and \textit{one} TTI. Depending on the packet size, channel quality and scheduling strategy, the transmission time $t_{tx}$ can vary from one to multiple TTIs. Considering the small payload of URLLC traffic, we assume $t_{tx} = 1$ TTI in this work. The processing time at the UE $(t_{up})$ is also assumed to be one TTI. 

In the case of failure to successfully decode the data message in the first transmission, a fast HARQ mechanism ensures quick retransmission of the message. In this case, the transmission is subject to additional delay(s), which includes the additional time it takes to transmit a negative acknowledgement (NACK), process it at the BS and schedule the packet for retransmission. The HARQ round trip time (RTT) $t_{HARQ}^{RTT}$, defined from the start time of the first transmission until the start time of the retransmission, is assumed to be \textit{four} TTIs.


\subsection{System Assumptions}

In order to meet the stringent latency target of URLLC services, a very flexible frame structure for 5G NR offering different options to shorten the TTI duration, as compared to LTE, is introduced by 3GPP~\cite{3gppTS38211}. In particular, the subcarrier spacing (SCS) can be expanded up to $480$ kHz (note, SCS of $480$ kHz is specified but not supported in release-15~\cite{3gppTR38912}), thereby reducing the minimum scheduling interval considerably. In addition, `mini-slots' are introduced whereby the number of orthogonal frequency-division multiplexed (OFDM) symbols per TTI can also vary. In contrast with the LTE slot duration of $14$ OFDM symbols per TTI, mini-slots in 5G NR can compose of $1 - 13$ symbols. The recommended mini-slot lengths are two, four and seven symbols, corresponding to TTIs of $0.07, 0.143$ and $0.25$ ms at $30$ kHz SCS. This allows shorter transmission slots without increasing the SCS, which is particularly suitable for low frequency bands. 

In this work, we assume a four symbol mini-slot at $30$ kHz SCS, resulting in a transmission duration of $0.143$ ms. This leaves sufficient time budget for the first transmission, processing at the UE and a single HARQ retransmission (if needed) within the one ms latency target for URLLC services, as illustrated in Figure~\ref{fig:processingTime}. In case of retransmission(s), multiple retransmitted packets are combined using Chase combining, resulting in a boost in the desired signal power~\cite{FPD01_chaseCombining}. 

In order to further enhance low latency support, we adopt the \textit{in-resource control signaling} proposed in~\cite{pedersen_5Gframe_comMag2016} along with front-loading of the demodulation reference symbol (DMRS)~\cite{mahmood_rrmEmbbMmtc_vtcFall2016}. The main idea is to embed control information on the fly at the start of time-frequency resources allocated to the user in the downlink. This allows the metadata to be processed and decoded as soon as it is received, i.e. while the data is still being received. From the latency perspective, this is advantageous as it allows performing channel estimation earlier and can enable early HARQ feedback as detailed in~\cite{berardinelli_earlyHARQfeedback_vtc2016}. HARQ feedbacks are always assumed to be received correctly.

We assume that the metadata (i.e., the control information needed to decode the transmission) and the data for the $l^{th}$ transmission are encoded separately with different target block error rates (BLER) given by $P^{m,l}_{e}$ and $P^{d,l}_{e}$, respectively; where $l \in \{1,2\}$. Upon Chase combining following a retransmission, the data outage probability is given by $P^{d,c}_{e}$. Note that, $P^{d,l}_{e} > P^{d,c}_{e} \, \forall l \in \{1,2\}$. Table ~\ref{tab:Pout_overview} provides an overview of the different outage probabilities introduced and derived in this contribution.

In terms of MC operation, we assume that the high priority URLLC packets are scheduled at the secondary node immediately upon arrival at the PDCP layer. Thus, transmission through each of the secondary nodes can also accommodate a single retransmission within the one ms latency budget, if needed. 

\begin{table}[htb]
\centering
\caption{Introduced and derived outage probabilities}
\label{tab:Pout_overview}
\renewcommand{\arraystretch}{1.3}
\begin{tabular}{l  p{6.5cm}}
\toprule 
\multicolumn{2}{l}{\textbf{\textit{Introduced Outage Probabilities:}}}\\
$P_{e}^{m,l}$ & Metadata BLER target in the $l^{th}$ transmission\\ 
$P_{e}^{d,l}$ & Data BLER target in the $l^{th}$ transmission\\ 
$P_{e}^{d,c}$ & Data error probability after Chase combining following retransmission\\ 
\midrule
\multicolumn{2}{l}{\textbf{\textit{Derived Outage Probabilities}}}\\
$P_{out}^{SC}$ & Outage probability of the baseline single-connectivity.\\
$P_{out}^{MC}$ & Outage probability of MC with data duplication.\\
\bottomrule
\end{tabular} 

\end{table}


\section{Reliability Enhancement with Multi-Connectivity}
\label{sec:reliabilitEnhancementWithMC}

In this section, we present an analytical derivation of the reliability enhancement with MC with PDCP duplication as defined in 3GPP standard~\cite{3gppTS37340}, and compare it against the baseline single-connectivity (SC) transmission. Ideal link adaptation is assumed, i.e. the achieved outage probabilities after transmission are assumed to be the same as the transmission BLER targets. 


\subsection{Baseline outage probability with Single-Connectivity}
\label{sec:SC_analysis}
The outage probability with SC considering separate BLER targets for the metadata and data is analyzed first. Due to its critical nature, we assume that the metadata is encoded with a lower BLER target, i.e. $P^{m,l}_{e} < P^{d,l}_{e}.$ 

Under conventional eMBB transmissions, the BLER target of the data part is much higher than that of the metadata (i.e., $P^{d,l}_{e} \gg P^{m,l}_{e}$). Hence, the impact of the metadata outage on the overall outage probability is negligible, and the outage probability can be readily approximated by the data BLER target $P^{d,l}_{e}.$ However, the same cannot be assumed for URLLC services requiring high reliability where the data is also transmitted with a stringent BLER target. 

\begin{figure}[thb]
    \centering
    \includegraphics[width=0.99\columnwidth]{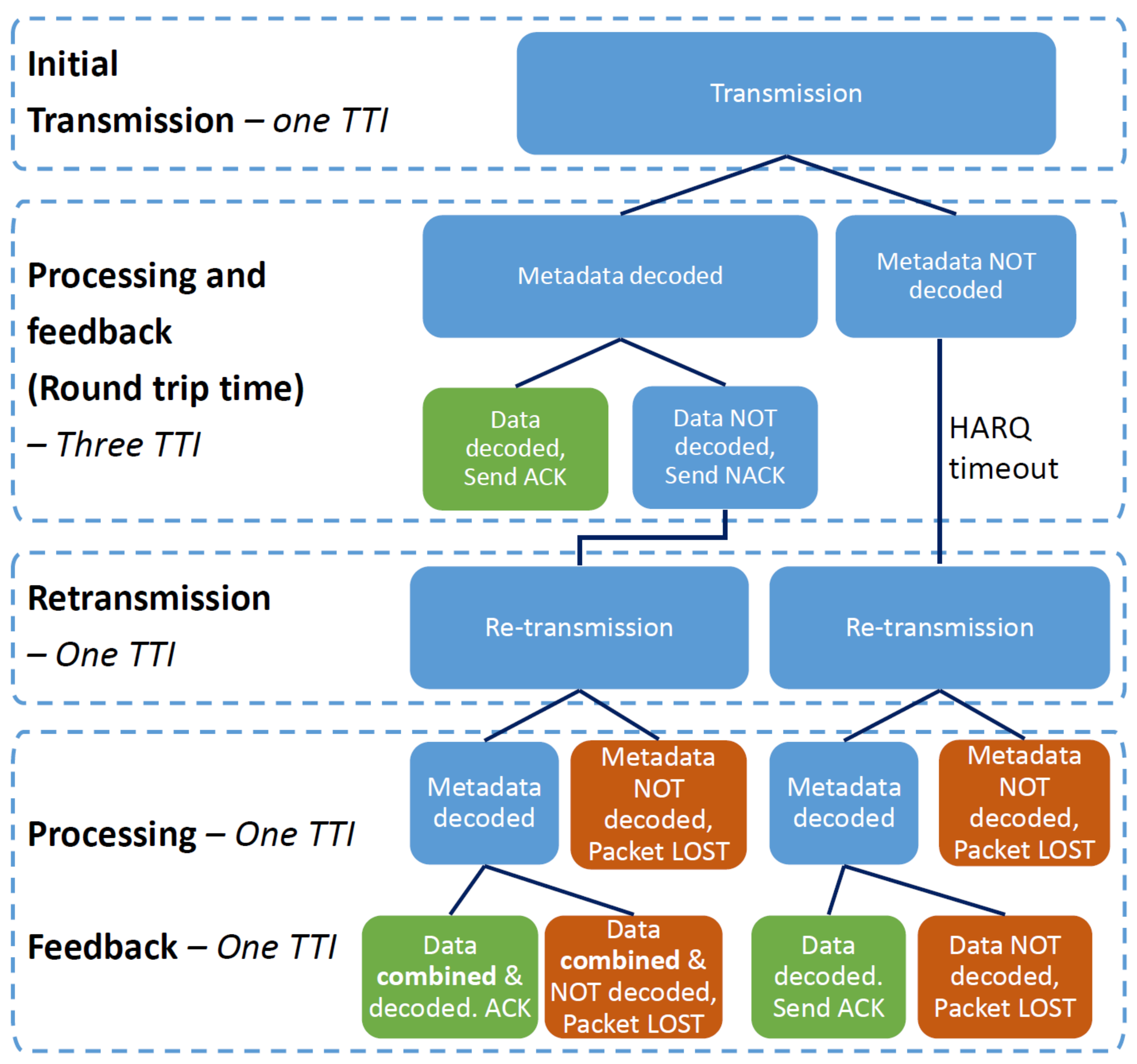}
    \caption{Difference possible reception events in the single-connectivity scenario.}
    \label{fig:txEvents_SC}
\end{figure}

The events that can occur upon transmission are depicted in Figure~\ref{fig:txEvents_SC}. There are three different possible outcomes of processing the first transmission at the receiver: failure to decode the metadata (with probability $P^{m,1}_{e}$), metadata decoded but failure to decode the data (with probability $\left(1 - P^{m,1}_{e} \right) P^{d,1}_{e}$) and successful decoding of the data packet in the first attempt. The probability of success in the first transmissions is given by
\begin{align}
\label{eq:succProb1}
P_{succ}^{SC,1} = \left(1 - P^{m,1}_{e}\right)\left(1 - P^{d,1}_{e}\right).
\end{align}

A HARQ NACK cannot be transmitted if the metadata is not successfully decoded in the first transmission. This leads to a HARQ \textit{time out}, which occurs when a HARQ feedback (ACK/NACK) is not received within a pre-defined time interval. The transmitter then retransmits the packet assuming the initial transmission failed. However, there is no possibility of Chase combining in this case since the control information needed to identify the packet in the first transmission was not successfully decoded. Thus, the success probability with retransmission following a HARQ time out is given by 
\begin{multline}
\label{eq:succProb2-TO}
P_{succ}^{SC,2-TO} = P^{m,1}_{e}\left(1 - P^{m,2}_{e}\right)\left(1 - P^{d,2}_{e}\right).
\end{multline}
In this study, we set the HARQ \textit{time out} time and the time it takes to process the retransmission at the BS $(t_{bp})$ to be equal to three TTIs, thus ensuring the same retransmission latency as that with HARQ retransmission. 

In the event of receiving a NACK, the packet is retransmitted. Reception of the retransmitted data can be attempted by Chase combining with the initially received data whose metadata was successfully decoded in the first transmission. In this case, the probability of successful decoding is $\mathrm{Pr}\left[\gamma_{c} \geq \gamma_{t} \right],$ where $\gamma_{c} = \gamma_{1} + \gamma_{2}$ is the achieved signal to interference plus noise ratio (SINR) following Chase combining. The SINR of the $l^{th}$ transmission, and the target SINR, are denoted by $\gamma_{l}$ and $\gamma_{t}$, respectively. The decoding success probability in this case of HARQ retransmission, $P_{succ}^{SC,2-NR}$, is given by
\begin{multline}
P_{succ}^{SC,2-NR} = \left(1 - P^{m,1}_{e} \right) P^{d,1}_{e} \left(1 - P^{m,2}_{e} \right) \times \\
\left( 1 - \mathrm{Pr}\left[\gamma_{c} < \gamma_{t} | \gamma_{1} < \gamma_{t} \right] \right).
\end{multline}
Using Baye's rule, $\mathrm{Pr}\left[\gamma_{c} < \gamma_{t} | \gamma_{1} < \gamma_{t} \right]$ can be expressed as ${}^{\mathrm{Pr}\left[\gamma_{c} < \gamma_{t}, \gamma_{1} < \gamma_{t} \right]}/_{\mathrm{Pr}\left[\gamma_{1} < \gamma_{t} \right]}$. Since $\gamma_{1} < \gamma_{c}$, we have $\mathrm{Pr}\left[\gamma_{c} < \gamma_{t}, \gamma_{1} < \gamma_{t} \right] = P^{d,c}_{e}$. Hence, $P_{succ}^{SC,2-NR}$ can be further simplified to 
\begin{multline}
\label{eq:succProb2-NR}
P_{succ}^{SC,2-NR} = (1 - P^{m,1}_{e} ) (1 - P^{m,2}_{e} ) (P^{d,1}_{e} - P^{d,c}_{e} ).
\end{multline}

The success probability following a retransmission is then equal to the sum of the success probability of retransmission after time-out, and that upon retransmission following a NACK. The sum can be expressed as
\begin{multline}
\label{eq:succ_SC_retx}
P_{succ}^{SC,2} = \left(1 - P^{m,2}_{e}\right)
\left[ P^{m,1}_{e} \left(1 - P^{d,2}_{e}\right)  + \right.\\
\left. \left(1 - P^{m,1}_{e}\right) \left(P^{d,1}_{e} - P^{d,c}_{e}\right) \right].
\end{multline}

Consequently, the total outage probability for the baseline SC scenario is 
\begin{equation}
\label{eq:outage_SC}
P_{out}^{SC} = 1 - P_{succ}^{SC,1} - P_{succ}^{SC,2}.
\end{equation}


\subsection{Outage Probability Analysis in Multi-Connectivity Scenario}
\label{sec:MC_analysis}
We now analyze the outage probability of MC considering data duplication at the PDCP layer, as defined in 3GPP release-15~\cite{3gppTS37340}. In this MC variant, data packets are duplicated and shared between the master node and the secondary node(s) at the PDCP layer. The packets are then transmitted independently from each node, i.e., they can have different MCS and transmitted over different resource blocks (RB). At the UE end, the lower layers up to the radio link control layer treat each of the packet received from the different nodes as separate packets and attempt to decode them individually. Successfully received packets are then forwarded to the PDCP layer. If multiple copies are successfully decoded, the PDCP layer keeps the first received packet while discarding any later copies. 

Assuming independent transmissions of the same data packet over $M$ nodes, the packet is lost if it is not successfully decoded from any of the $M$ nodes. Hence, the outage probability is given by 
\begin{equation}
\label{eq:outage_MC}
P_{out}^{MC}(M) = \prod_{n=1}^{M} P_{out,n}^{SC},
\end{equation}
where $P_{out,n}^{SC}$ is the outage probability through the $n^{th}$ node, and can be evaluated using Eq.~\eqref{eq:outage_SC}. In the case of identical outage probabilities through all links (i.e., transmission with the same BLER targets from all nodes), the outage probability further simplifies to $P_{out}^{MC}(M) = \left(P_{out}^{SC}\right)^{M}.$


\section{Resource Usage Analysis}
\label{sec:resourceUsage}

This section evaluates the resource usage of the baseline SC and MC with PDCP duplication using results from finite blocklength theory~\cite{polyanskiy_trIT2010}. The number of information bits $L$ that can be transmitted with decoding error probability $P_{e}$ in $R$ channel use in an additive white Gaussian noise (AWGN) channel with a given SINR $\gamma$ is
\begin{align}
	\label{eq:polyanskiy}
	L = R C(\gamma) - Q^{-1}(P_e)\sqrt{R V(\gamma)} + \mathcal{O}(\log_2 R),
\end{align}
where $C(\gamma) = \log_2 (1 + \gamma)$ is the Shannon capacity of AWGN channels under infinite blocklength regime, $V(\gamma) = \frac{1}{\ln(2)^2} \left( 1 - \frac{1}{\left( 1 + \gamma \right)^2}\right)$ is the channel dispersion (measured in squared information units per channel use) and $Q^{-1}(\cdot)$ is the inverse of the Q-function. Using the above, the channel usage $R$ can be approximated as~\cite{AV_jsac2018}
\begin{multline}
	\label{eq:channelUsageAnand}
	R \approx \frac{L}{C(\gamma)} + \frac{Q^{-1}(P_e)^2 V(\gamma)}{2 C(\gamma)^2} \times \\
	\left[ 1 + \sqrt{1 + \frac{4 L C(\gamma)}{Q^{-1}(P_e)^2 V(\gamma)}}\right].	
\end{multline}

For a given transmission schemes $\Pi \in \{SC, MC\}$, we first evaluate the BLER target $P_{e}^{\Pi}$ that can achieve the desired $10^{-5}$ URLLC outage probability using the equations derived in Sections~\ref{sec:SC_analysis} and~\ref{sec:MC_analysis}. The channel use per single transmission, $R_{\Pi}$, is then calculated by inserting the corresponding values of $P_{e}^{\Pi}$ into Eq.~\eqref{eq:channelUsageAnand}. Finally, the total resource usage including the effect of retransmission is evaluated henceforth. 

\subsubsection{Single-Connectivity}
For single connectivity, the resource usage is $R_{SC}$ with probability $P_{succ}^{SC,1}$ and $2R_{SC}$ with probability $1 - P_{succ}^{SC,1}.$ Hence, the total resource usage, $\mathcal{U}_{SC}$, is straightforwardly obtained as 
\begin{align}
	\label{eq:ru_sc}
	\mathcal{U}_{SC} = \left(2 - P_{succ}^{SC,1}\right) R_{SC}.
\end{align}

\subsubsection{Multi-Connectivity} 
For MC with PDCP duplication through $M$ nodes, each transmissions are independent with a retransmission occurring in the case of failure of that transmission. Hence, the transmission (or retransmission) through a given node is not cancelled even if the packet has already been correctly decoded from the transmission through other nodes. Thus, $M R_{MC}$ channel uses are used if the initial transmissions through all $M$ nodes are successful, $(M + 1) R_{MC}$ channel uses are used if one initial transmission fails, and so on. (Here, we assume, for the ease of presentation, the achieved SINR through all $M$ nodes are the same. This can happen, e.g., when considering a UE at the cell edge with equal received power from multiple BSs. In general, the channel use for each node can be obtained by inserting the appropriate SINR value in Equation~\eqref{eq:channelUsageAnand}. The corresponding resource usage can then be calculated easily using any numerical computing software.)

In other words, the channel usage is a random variable that can take the values $(M + n)R_{MC}$, for $n = 0 \ldots M$, with probability $\binom{M}{n} \left(P_{succ}^{SC,1}\right)^{M-n} \left(1 - P_{succ}^{SC,1}\right)^{n}$. The total resource usage can then be calculated by summing the above for $n = 0 \ldots M.$ After some algebraic manipulation, this yields 
\begin{align}
	\label{eq:ru_mc}
	\mathcal{U}_{MC} = M \left(2 - P_{succ}^{SC,1}\right) R_{MC}.
\end{align}


\section{Numerical Results}
\label{sec:results} 

This section presents numerical validation of the derived analytical results. For simplicity, we assume that the outage probabilities remain unchanged over the initial transmission and the retransmission, i.e. $P^{d,1}_{e} = P^{d,2}_{e} = P^{d}_{e}$ and $ P^{m,1}_{e} = P^{m,2}_{e} = P^{m}_{e}$. This is a reasonable assumption as the time between retransmissions is very short. Moreover, the assumed payload and metadata size is $32$ and $16$ Bytes, respectively. 

\subsection{Outage Probability as a Function of $P_e^{d}$}
\begin{figure}[htb]
    \centering
    \includegraphics[width=0.99\columnwidth]{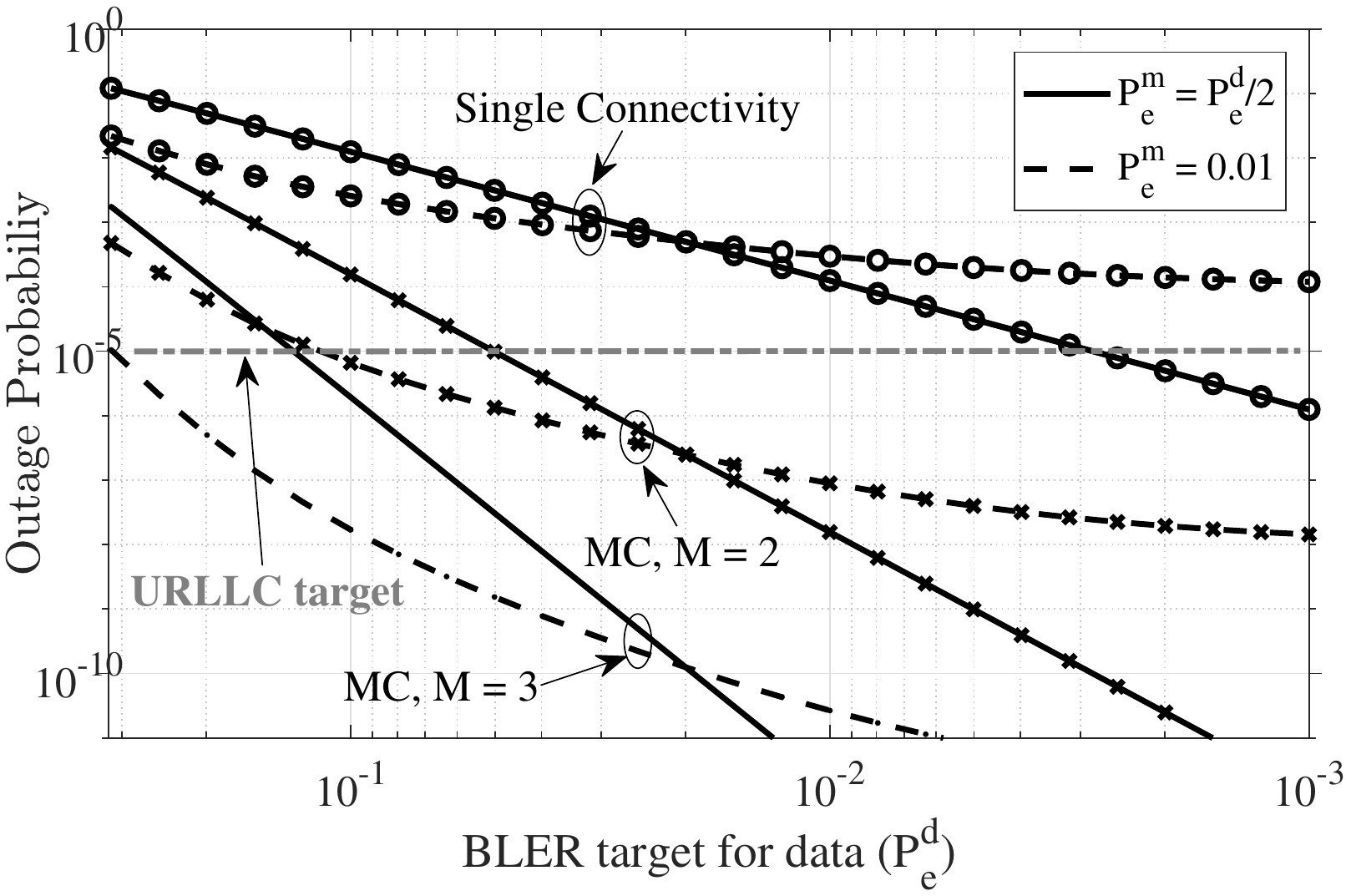}
    \caption{Outage Probabilities with single-connectivity and multi-connectivity transmission for $P_e^{m} = [{}^{P_e^{d}}/_{2}, 0.01]$ and $M = [2, 3].$}
    \label{fig:outageProb}
\end{figure}

The derived outage probabilities with SC and MC are presented as a function of the BLER target on the data channel $\left( P_e^{d} \right)$ in Figure~\ref{fig:outageProb}. Two different metadata BLER targets are considered, namely $P_e^{m} = {}^{P_e^{d}}/_{2}$ and $P_e^{m} = 1\%$. MC results are evaluated with $M = 2$ and $M = 3$. 

With SC, the outage probability remains above the URLLC target of $10^{-5}$ even for $P_e^{d}$ as low as $1\%.$ In fact, $10^{-5}$ outage is only achieved with $P_e^{m}$ and $P_e^{d}$ at about $0.15\%$ and $0.3 \%$, respectively. However, the targeted reliability can be met at much higher BLER targets with MC transmission, even when only two nodes are involved (i.e., $M=2$). 

Comparing the outage probabilities, we can observe a large gap between the performance of SC and the MC schemes. This indicates that there is a clear advantage in terms of the outage probability in transmitting multiple copies of the packet, especially at the levels targeted for URLLC applications. Since decoding the metadata is more critical than the data itself, we observe clear advantage in having a lower BLER target for the metadata. 


\subsection{Resource Usage Analysis}

\begin{figure}[thb]
    \centering
    \includegraphics[width=0.99\columnwidth]{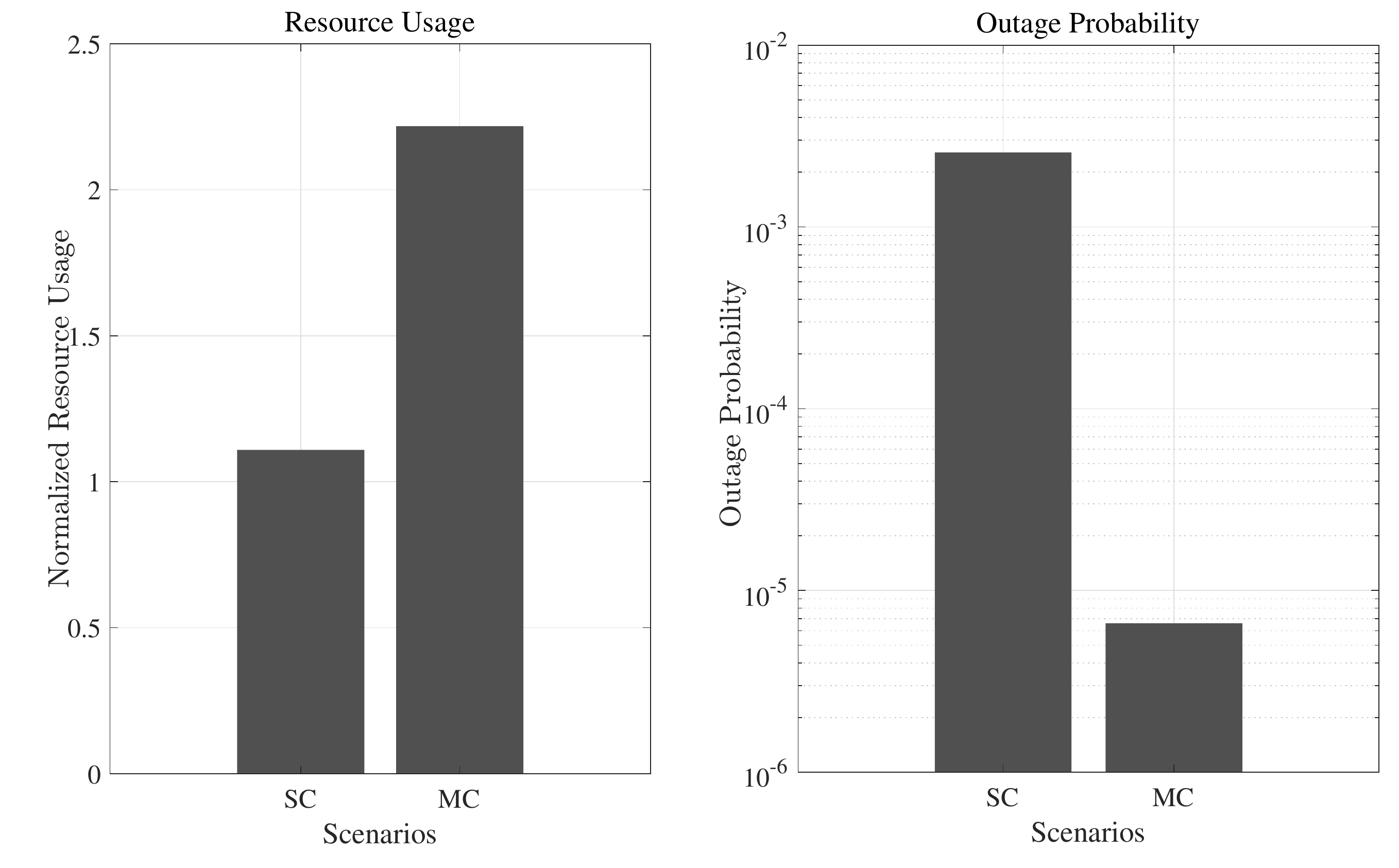}
    \caption{Resource usage and corresponding outage probabilities with SC and MC, $P_e^{m} = 1\%, P_e^{d} = 10\%$ and $M = 2.$}
    \label{fig:resourceUsage}
\end{figure}

Figure~\ref{fig:resourceUsage} presents the resource usage and corresponding outage probabilities for SC and MC transmission scheme with $M = 2$, where the BLER targets for metadata and data are fixed at $1\%$ and $10\%$ respectively. Since the same BLER targets are assumed for all schemes, the resource usage is normalized by the resource utilization for a single transmission. Please note that we assume the same SINR is achieved through the master and the secondary node, as discussed earlier. This is only to facilitate the analytical derivations and obtain meaningful insights into the performance trend. Performance results with different SINRs can easily be evaluated numerically.

Single-connectivity is expectedly the most resource efficient, though this comes at an outage probability that is several orders of magnitude higher than the MC scheme. Thus, the price to pay for the higher reliability with MC is the almost doubling of the resource usage and additional signalling overhead. Note however that, resource efficiency is not the main performance indicator in many applications requiring high reliability. Nonetheless, this provides a strong motivation for investigating more resource efficient MC schemes.

In the rest of this section, we calculate the resource utilization at $1- 10^{-5}$ reliability target using the results derived in Section~\ref{sec:resourceUsage}. Table~\ref{tab:Pout_at_URLLCtarget} shows the required BLER targets for achieving $1- 10^{-5}$ reliability target for SC and MC transmission scheme with $M = 2$, assuming $P_e^{m} = P_e^{d}.$ The corresponding channel use per transmission at an SINR of $10$ dB derived using results from finite blocklength theory and the final resource usage with retransmission are also listed. We observe that the required BLER target with SC is more than an order of magnitude lower than that with MC. 

\begin{figure}[htb]
    \centering
    \includegraphics[width=0.99\columnwidth]{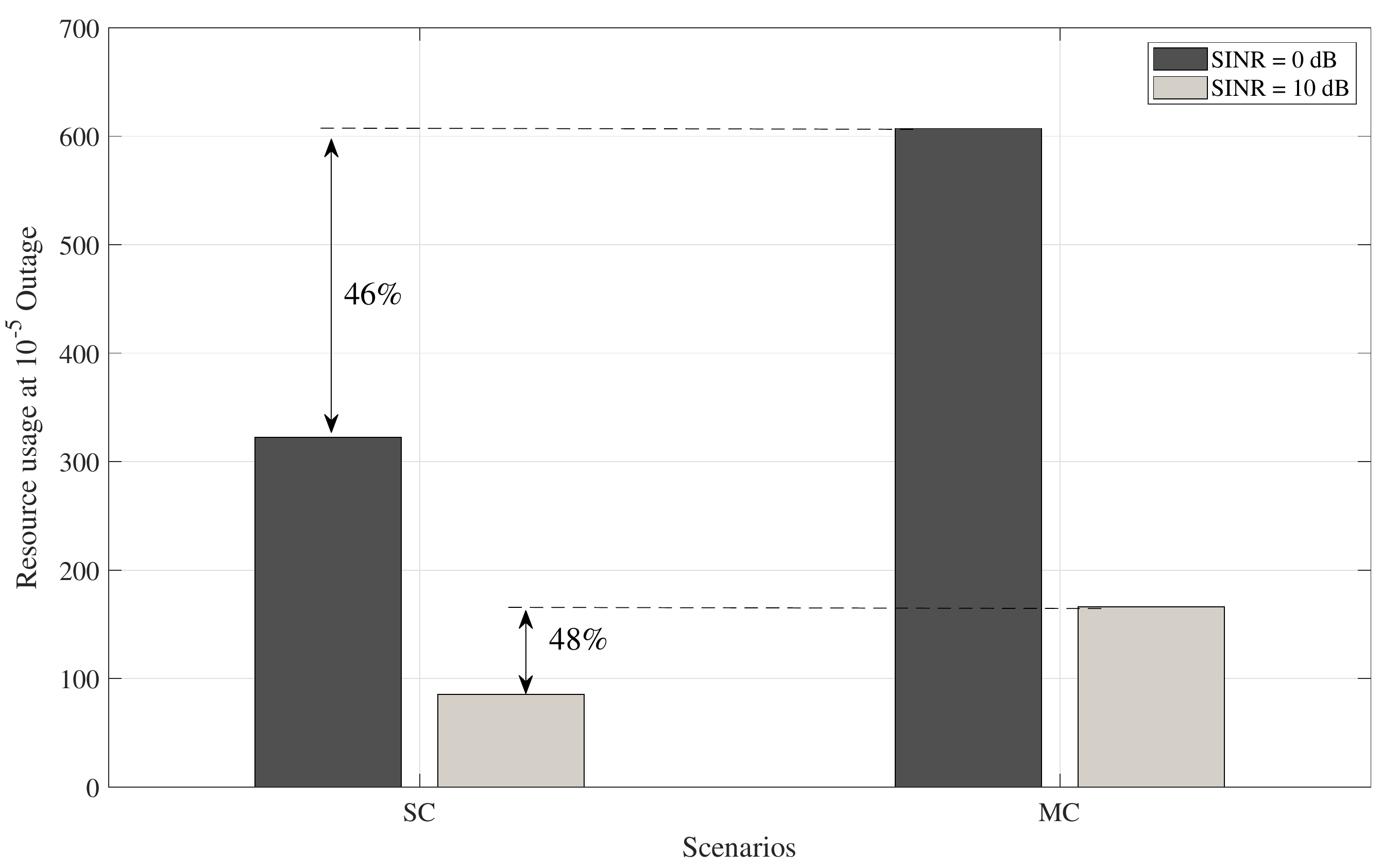}
    \caption{Resource usage of the single-connectivity and multi-connectivity scheme at $1- 10^{-5}$ reliability.}
    \label{fig:resourceUtilAtURLLC}
\end{figure}

Figure~\ref{fig:resourceUtilAtURLLC} presents the total resource usage in terms of the channel use with a BLER target set to achieve $10^{-5}$ outage probability for SINRs of $0$ and $10$ dB. Even after taking into account the lower BLER targets required for achieving $10^{-5}$ outage probability, SC is found to be more resource efficient compared to the considered MC scheme. In fact, $46\%$ to $48\%$ less resources are required with SC, depending on the SINR value. However, the resourced required to achieve a given reliability target with SC may not always be available at a given node. Furthermore, the success probability following the single transmission is much higher with MC, meaning that it has clear advantages in applications with a tight latency budget where even a single retransmission cannot be accommodated~\cite{BML+18_wirt}. 


\begin{table}[thb]
\centering
\caption{Resource usage at $1- 10^{-5}$ reliability target}
\label{tab:Pout_at_URLLCtarget}
\renewcommand{\arraystretch}{1.3}
\begin{tabular}{p{3cm} c c c }
\textit{Tx. scheme $(\Pi)$} & \textit{BLER target} & $R_{\Pi}$ & $\mathcal{U}_{\Pi}$ \\
\hline 
Single-Connectivity& $0.183\%$ & $85.14$ & $85.44$\\
Multi-Connectivity& $3.28\%$ & $80.88$ & $166.12$\\
\end{tabular} 

\end{table}

\section{Conclusions}
\label{sec:conclusion}

Multi-connectivity is proposed as a potential reliability enhancement solution for URLLC applications. The outage probability considering MC with PDCP duplication as defined by 3GPP is derived and compared against baseline SC scheme in this paper. In contrast with existing works, the reliability of the control channel is specifically considered in the outage probability evaluation. The corresponding resource usages are also derived. Collectively, the derived outage probability and resource usage analysis allow comparing the cost-performance trade-offs of MC as a reliability enhancement solution for URLLC services.

The obtained analytical results show that MC can greatly enhance the outage probability at the expense of increased resource usage. In particular, the outage probability is enhanced by several orders of magnitude, at the expense of almost doubling of the resource usage. From a resource utilization perspective, single-connectivity is more resource efficient. However, MC is more desirable from a reliability aspect since the reliability levels targeted for URLLC applications may not always be possible with SC. Furthermore, it has clear advantages in scenarios where even a single retransmission cannot be accommodated, for example in certain industrial use cases with less than one ms latency requirements. Our future work includes investigating more resource efficient MC transmission schemes.



\end{document}